\documentstyle[twoside,fleqn,espcrc2,epsfig]{article}
\title{ Spinodal Decomposition and the Deconfining Phase 
Transition\thanks{This work was in part supported by the US 
Department of Energy under contract DE-FG02-97ER41022.} }
\author{Bernd A. Berg\address{Department of Physics, 
Florida State University, Tallahassee, FL~32306, USA}$^,$\address{
School of Computational Science and Information Technology,\\
~~~Florida State University, Tallahassee, FL~32306, USA},
Urs M. Heller\address{American Physical Society, One Research Road,
Box 9000, Ridge, NY 11961, USA},
Hildegard Meyer-Ortmanns\address{ School of Engineering and Science, 
International University Bremen,\\ 
~~~P.O. Box 750561, D-28725 Bremen, Germany},
and Alexander Velytsky$^{\rm a, b}$
}
\begin{document}
\begin{abstract}
We study the Glauber dynamics of simple spin systems to identify
dynamical scenarios which may be of relevance for the deconfining 
phase transition in heavy ion collisions.
\end{abstract}
\date{\today}
\maketitle

\section{Introduction}

Lattice gauge theory investigations of the deconfining phase 
transitions have mainly been limited to equilibrium studies (an 
exception is the work by Miller and Ogilvie~\cite{MiOg02}). 
The equilibrium transition with two massless quarks is likely
second order and it becomes a crossover with two light quarks
(and the heavier strange quark)~\cite{Bielefeld}.
In nature the finite temperature phase transition is governed 
by temperature driven dynamics.

Early universe: We have a slow cooling process ($10^{-5}-10^{-6} >> 
10^{-23}\,$s). Most likely, the effects of the dynamics are negligible 
and no signals of the transition are observable nowadays.

Heavy ion collisions -- Bjorken's~\cite{Bj83} standard scenario:
In the center of mass frame the incident nuclei are Lorentz contracted
into pancake shapes. They pass through each other and leave behind a
region of hot vacuum. The heating is presumably not slow on the 
relaxation time scale, but usually considered as a quench, i.e., an 
instantaneous process. Subsequently, the cooling is not much slower
than the scale of $10^{-23}\,$s. 


Quenching is a process in which the temperature in the symmetric 
phase below $T_c$ is raised instantaneously to a temperature in the 
broken phase above $T_c$. Quenching has been much studied in condensed 
matter physics. One finds that the dynamics of long-wave modes groups 
theories into dynamical universality classes described by the same 
equations of motion. 

Correlated domains emerge and grow with time in such a way that 
the correlation function of a generic field $\phi$ has the simple 
scaling form 
\begin{eqnarray}
 g(\vec{r},\vec{r}{\,'},t) & = &
\langle \phi (\vec{r},t) \phi(\vec{r}{\,'},t) \rangle \\ 
& = & f(|\vec{r}-\vec{r}{\,'}|/L(t)),\ L(t)\sim t^p \nonumber
\end{eqnarray}
where the exponent $p$ depends only on the dynamical universality 
class. Properties of the Fourier transforms, called
structure functions or structure factors, allow to differentiate 
between the scenarios of nucleation and spinodal
decomposition. For spinodal decomposition clusters grow on every 
length scale and one finds pronounced maxima in the low-momentum 
structure functions. Nucleation is dominated by the growth of the 
largest clusters and the maxima of the low-momentum structure 
functions are insignificant.

In real heavy ion collisions, the phase transition will probably 
be in between a slow equilibrium process and a quench. Therefore, we 
focus on the dynamics of hysteresis loops. We supplement 
the results using equilibrium and quenching data.

We are interested in the response of an ensemble of Polyakov loops 
to a change of temperature from the symmetric to the broken phase. 
Glauber dynamics is our best shot which allows us to study the
influence of the speed of the heating or cooling process.
It includes canonical Metropolis and heat bath updating 
procedures, which imitate local fluctuations of nature. 
The aim of such a study is not to make precise quantitative 
predictions, but to identify possible dynamical scenarios.

A direct lattice QCD simulation is very CPU time intensive. Therefore,
the 3d 3-state Potts model serves in our simulations as an effective 
spin model, which is supposed to be in the same universality 
class~\cite{SY82}. A magnetic field allows to imitate the effect 
of finite quark masses~\cite{BU83}.

In our first step, reported here, the 2d q-state Potts model
serves as a test laboratory. The advantage is that some results are
analytically known.
We simulate the 2d 2-, 4-, 5- and 10-state Potts models,
corresponding to a weak 2nd order ($\alpha=0$), a "strong" 2nd order 
($\alpha = 2/3$), a weak first order ($\alpha =1, \triangle e_l=0.03$) 
and a strong first order ($\alpha =1, \triangle e_l=0.35$) phase 
transition, where $\alpha$ is the critical exponent of the specific 
heat and $\triangle e_l$ the latent heat per link.  We study the 
hysteresis with change of coupling
\begin{equation} \label{hysteresis}
\triangle\beta={2(\beta_{\max}-\beta_{\min})\over n_{\beta}\,L^2}\ . 
\end{equation}
This is a dynamics which slows down with increasing lattice size. 
At least 640 cycles are performed per lattice size. This gives an
ensemble of non-equilibrium configurations. The equilibrium is 
recovered for $n_{\beta}\to\infty$.

In the following results are reported for the internal energy $E$ 
as an example of a bulk quantity, stochastic clusters and
structure functions. 

\section{Results}

\begin{figure}[ht] \vspace{-8mm} \begin{center}
\epsfig{figure=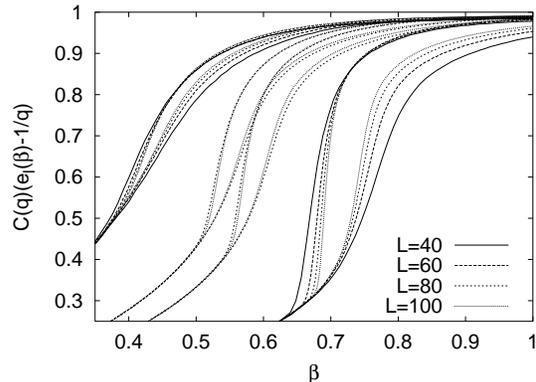,width=\columnwidth} \vspace{-15mm}
\caption{ Energy hysteresis loops. From left to right for 
$q=2,\,4,\,5$ and 10.}
\label{fig_hys_e} \end{center} \vspace{-8mm}
\end{figure}

For a properly rescaled internal energy hysteresis cycles for our
$n_{\beta}=1$ dynamics are shown in Fig.\ref{fig_hys_e}. The
openings are finite volume estimators of the latent heat.
Performing infinite volume extrapolations to the latent heat, 
the resulting fits are only accurate for the $q=10$ strong first 
order phase transitions.  Otherwise, the results are too high when
compared with equilibrium values, resulting in a `dynamical' latent 
heat for the second order transition.

\begin{figure}[ht] \vspace{-8mm} \begin{center}
\epsfig{figure=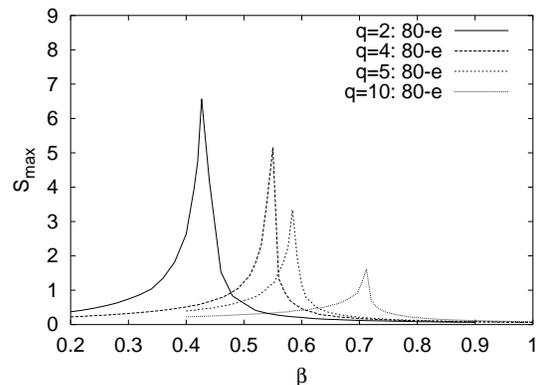,width=\columnwidth} \vspace{-15mm}
\caption{ Equilibrium results for the largest cluster surface. }
\label{fig_clsm_eq} \end{center} \vspace{-10mm}
\end{figure}

\begin{figure}[ht] \vspace{-8mm} \begin{center}
\epsfig{figure=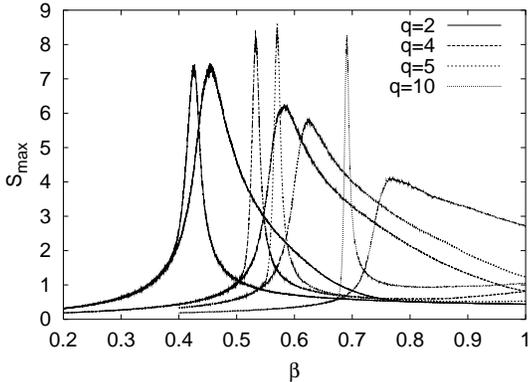,width=\columnwidth} \vspace{-15mm}
\caption{ Dynamical $n_{\beta}=1$ results for the largest cluster surface. }
\label{fig_clsm_1V} \end{center} \vspace{-10mm}
\end{figure}

Fortuin and Kasteleyn introduced stochastic clusters, which allow to 
formulate the Potts models as cluster systems. For a fixed configuration 
the clusters are build as in the Swendsen-Wang algorithm. We have 
studied the Glauber dynamics of many cluster properties and report here 
the behavior of the largest cluster surface, which has pronounced peaks 
close to the phase transition. Fig.\ref{fig_clsm_eq} shows decreasing 
peak heights for increasing $q$. This is expected for a scenario,
which moves from spinodal decomposition to nucleation. 
Fig.\ref{fig_clsm_1V} suggests than that for our $n_{\beta}=1$
dynamics the transition becomes always spinodal.

With $m_{q_0}=\langle\delta_{\sigma(\vec{r},t),q_0}\rangle$
the structure functions are defined by
\begin{eqnarray} \nonumber
 S(\vec{k},t)&=&\frac1{N_s^2}\sum_{q_0=0}^{q-1}
 \left\langle\left|\sum_{\vec{r}}\delta_{\sigma(\vec{r},t),q_0}
 \exp[i\vec{k}\vec{r}]\right|^2\right\rangle \\
 \label{structure_f} &-&\delta_{\vec{k},0} \sum_{q_0} m_{q_0}^2\ .
\end{eqnarray}
We recorded the low-lying momenta $k_i=2\pi L^{-1} n_i$ with
$n_1=(1,0)$ and $(0,1)$, $n_2=(1,1)$, $n_3=(2,0)$ and $(0,2)$,
$n_4=(2,1)$ and $(1,2)$, $n_5=(2,2)$.

Spinodal decomposition is characterized by an explosive growth 
in the low momentum modes, while the high momentum modes relax 
to their equilibrium values. Miller and Ogilvie~\cite{MiOg02} 
investigated structure functions under quenching for SU(2) and 
SU(3) lattice gauge theory. Here, we study the structure functions 
during hysteresis loops and under quenching. 

\begin{figure}[ht] \vspace{-8mm} \begin{center}
\epsfig{figure=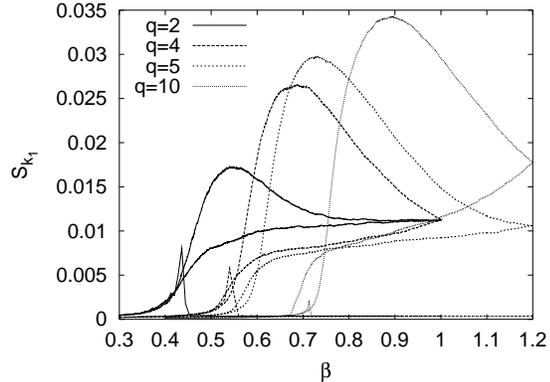,width=\columnwidth} \vspace{-15mm}
\caption{Equilibrium (small peaks) and $n_{\beta}=1$ dynamics
behavior of the $S_{k_1}(\beta)$ structure function.}
\label{fig_sf1k} \end{center} \vspace{-9mm}
\end{figure}

\begin{figure}[ht] \vspace{-8mm} \begin{center}
\epsfig{figure=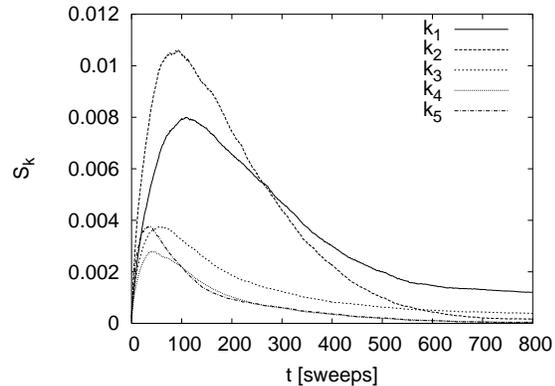,width=\columnwidth} \vspace{-15mm}
\caption{Time evolution for $q=2$ and $h=0.01$ after quenching 
$\beta$ from 0.2 to 0.6. }
\label{fig_q02h} \end{center} \vspace{-12mm}
\end{figure}

Fig.\ref{fig_sf1k} shows $S_{k_1}(\beta)$ hysteresis for our 
$n_{\beta}=1$ dynamics and confirms that this dynamics leads 
for all $q$ values to spinodal decomposition. In Fig.\ref{fig_q02h}
the time evolution of several structure functions after quenching 
is depicted for the $q=2$ model with an external magnetic field,
so that the transition is a crossover.

\section{Summary and Conclusions}

Energy hysteresis cycles allow to locate the temperature of the
equilibrium transition. For weak first order as well as second order
transitions the dynamics generates a rather large `dynamical' latent 
heat.

From our analysis of the Fortuin-Kasteleyn clusters and the structure 
functions we conclude: The dynamical transition shows always 
spinodal decomposition and the difference between 1st and 2nd 
order is washed out.

Signals of the spinodal decomposition may still survive when there 
is no proper phase transition, but a fast crossover. If a dynamics 
generates regions of misaligned Polyakov loops, one expects an 
enhanced production of low-energy gluons.

\end{document}